# Influence of Periodic Surface Nanopatterning Profiles on Series Resistance in Thin-Film Crystalline Silicon Heterojunction Solar Cells

Islam Abdo[1,2,3*], Christos Trompoukis[1,4], Loic Tous[1], Valérie Depauw[1], Rafik Guindi[3], Ivan Gordon[1], Ounsi El Daif[5]

[1] IMEC, Kapeldreef 75, 3001 Leuven, Belgium
[2] KACST-Intel Consortium Center of Excellence in Nano-manufacturing Applications (CENA), Riyadh, KSA
[3] Microelectronics System Design department, Nile University, Cairo, Egypt
[4] Katholieke Universiteit Leuven, Leuven, Belgium
[5] Qatar Environment and Energy Research Institute (QEERI), Qatar Foundation, Doha, Qatar

*Corresponding Author, islaam.abdo@gmail.com

ABSTRACT

In the frame of the development of thin crystalline silicon solar cell technologies, surface nanopatterning of silicon is gaining importance. Its impact on the material quality is however not yet fully controlled. We investigate here the influence of surface nanotexturing on the series resistance of a contacting scheme relevant for thin-film crystalline silicon heterojunction solar cells. Two-dimensional periodic nanotextures are fabricated using a combination of nanoimprint lithography and either dry or wet etching, while random pyramid texturing is used for benchmarking. We compare these texturing techniques in terms of their effect on the series resistance of a solar cell, through a study of the sheet resistance (Rsh) and contact resistance (Rc) of its front layers i.e. a sputtered transparent conductive oxide and evaporated metal contacts. We have found by four-point probe and the transfer length methods that dry-etched nanopatterns render the highest Rsh and Rc values. Wet-etched nanopatterns, on the other hand, have less impact on Rc and render Rsh similar to that obtained from the nontextured case.

## I. INTRODUCTION

Following the course of absorber layer thinning that can be of interest for cost reduction, increased open circuit voltage and even material flexibility, ultra-thin monocrystalline silicon solar cells with thicknesses in the range of 1-40 µm are considered. For such thin crystalline silicon (c-Si) layers, light trapping in the red part of the spectrum becomes crucial, and the high material consumption caused by random-pyramid texturing (RPT) cannot be tolerated. Thanks to advancements in nanophotonics, several light management techniques have emerged for thin-film solar cells allowing for light absorption enhancement with the lowest possible material consumption [1]–[8].

Among those surface texturing techniques, two-dimensional (2D) periodic surface nanopatterning [9][10] has been shown to improve light absorption into the photoactive layer of thin-film c-Si Heterojunction (HJ) solar cells, boosting their short-circuit current [11]–[14]. In such cells, surface nanopatterning has been integrated by the combination of nanoimprint lithography (NIL) [15] and reactive ion etching (RIE) producing what we call dry-NIL pattern. However, such light patterning technique has the downside of deteriorating the electrical performance of their host c-Si HJ solar cells. As we previously described in [11], electrical performance deterioration due to dry-NIL integration is twofold. Firstly, it reduces the open circuit voltage ($V_{OC}$) due to the profound surface damage caused by RIE, which in turn results in high surface recombination velocity and low minority-carrier life time [16]. Secondly, it reduces the solar cell fill factor (FF) due to poor contact between front layers that results in increased series resistance.

In order to tackle the first issue, a combination of

  

lithography with wet etching (wet-NIL) has been proposed as a variation of dry-NIL [16][17]. This combination enabled the fabrication of inverted nanopyramid structures that avoid the surface damage caused by RIE, and a nanotextured surface that can easily be passivated leading to lower surface recombination velocity and higher lifetime, without sacrificing the good optical propertied achieved by dry-NIL. As shown in [16], surface recombination velocities in the same order of magnitude as for RPT can be achieved with a-Si:H as surface passivation.

In this work we extend the study we performed earlier in [10] where we characterized the impact of dry-NIL on c-Si HJ solar cells. Herein we investigate the impact of surface nanotexturing on the series resistance, which can be characterized by breaking it down into sheet resistance ($R_{sh}$) and contact resistance ($R_c$) of the front layers that are affected by surface texturing i.e. transparent conductive oxide (TCO) and metal contacts. We measure $R_{sh}$ and $R_c$ for separate layers deposited on top of our textures of interest, dry-NIL, wet-NIL and RPT. For the sake of process simplicity, the textures were performed on wafers instead of thin c-Si films, as described in section II.B. $R_{sh}$ and $R_c$ are measured using the four-point-probe (4PP) method and the transfer length method (TLM), respectively.

## II. EXPERIMENTAL DETAILS

### A. Resistance Characterization Techniques

For measuring $R_{sh}$ and $R_c$, the TLM is used [18]. Typically, the TLM test structure, sometimes called the *ladder* structure [19], is composed of equally-spaced contacts of length *L*, width *W* and separation *d*, as depicted in Fig. 1(a). A bias is applied between a reference contact and another one separated by a distance *d*, allowing current to flow between them, from which measurement the resistance is calculated. The current measurement is repeated between the same reference contact and another one separated by a distance $iL + (i + 1)d$, where $i = 1,2,3, ....$ This enables measuring the resistance between contacts at different separation distances. The resulting curve is, typically, a straight line whose slope is $R_{sh}$ of the layer beneath the contacts, and whose intersection with the resistance axis is twice $R_c$. Both quantities, $R_c$ and $R_{sh}$, can be related by the following equation:

$$R_c = \frac{\rho_C}{L_T W}; \quad L_T = \sqrt{\rho_C/R_{sh}} \qquad (1)$$

Where $\rho_C$, $L_T$ and *W* are the specific contact resistivity, the transfer length and the contact width, respectively.

A variant of the *ladder* structure is the unequally spaced TLM structure with the voltage applied between adjacent contacts (see Fig. 1(b)). This structure helps evading current flowing into intermediate contacts of the ladder structure which results in erroneous measurements [19]. In section III.B.1 we discuss and compare both TLM structures in terms of their compatibility to our samples and the materials to be measured.

To characterize the impact of surface texturing on each layer individually we also use 4PP [20][21]. For accurate 4PP measurements, the sample under test should be sufficiently large to avoid sample size sensitivity during measurements. In our case we use 5 cm x 5 cm samples, a size that guarantees the validity of our measurements [21]. Moreover, for both 4PP and TLM, the sample under test should consist of the front layer to be measured placed on a nonconductive substrate. The nonconductive substrate is necessary to ensure that the applied current is only flowing in the front layer, avoiding current leakage.

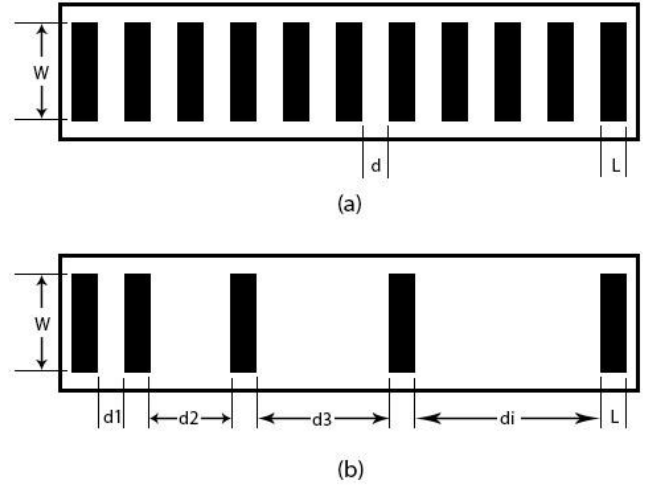

Fig. 1. TLM test structures with (a) equally-spaced, and (b) unequally-spaced contacts. The contacts dimensions were L=1mm, W=1cm, d=d1=1mm, d2=3cm, d3=5cm, …

### B. Sample Preparation

The fabrication of our test structures started by surface texturing of a supporting substrate on which the layers under test were deposited. The choice of the supporting substrate material is based on several aspects that ensure the relevance and accuracy of our measurements. Firstly, it should be resistive to avoid current leaking into it during measurements. Secondly, it should be sufficiently easy to etch during surface texturing, either by RPT, dry-NIL or wet-NIL. We chose the supporting substrate to be a lightly doped (<$10^{15}$ B atoms cm$^{-3}$) p-type c-Si mirror-polished Cz wafer, on which texturing is conducted, insulated by a 20-nm layer of intrinsic amorphous silicon (i/a-Si:H) deposited by a plasma enhanced chemical vapor deposition (PECVD), by the combination of $SiH_4$ (25 sccm) and $H_2$ (75 sccm) gasses at 250 ℃ for 20 seconds, using Plasmalab System 100 from Oxford Instruments.

We fabricated 2D periodic surface nanotextures using thermal NIL combined with either plasma etching in the case of dry-NIL [11] or wet chemical etching in the case of wet-NIL [16]. A thermoplastic resist was first spin-coated on then c-Si substrate. Then the pattern, with a predefined period of 900 nm, was transferred to the thermoplastic resist using an elastomer stamp and a heated hydraulic press. Etching was then used to transfer the pattern from the thermal resist, which acts as an etch mask, to the slab. In the case of dry-NIL, RIE was applied with the appropriate power, pressure, gas ($SF_6$ and $O_2$) concentration and time to etch silicon for the desired shape and depth of the nanostructures. RIE was done using a capacitively coupled plasma etcher, ML200 from DSE.

As for wet-NIL, a solution of 10% diluted tetramethyl ammonium hydroxide (TMAH) was used as an etchant. However, the resist is not adequately resistant to the TMAH solution. We therefore used a 100-nm SiO$_x$ layer deposited by PECVD as an intermediate. As depicted in Fig. 2, after the pattern transfer to the thermoplastic resist, we transferred the pattern from the resist mask to the SiO$_x$ mask using dry etching. TMAH was then used to transfer the pattern from the SiO$_x$ mask to the c-Si slab. The SiO$_x$ mask was then removed by 1% diluted HF solution.

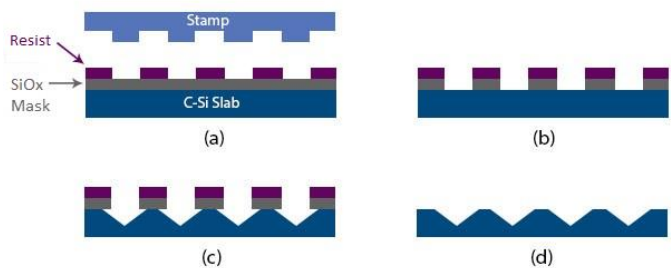

Fig. 2. Wet-NIL process: (a) pattern transfer from the stamp to the resist. (b) SiO$_x$ dry etching using CHF$_3$. (c) TMAH etching of c-Si forming inverted pyramids. (c) Final wet-NIL pattern after removal of both thermal-resist and SiO$_x$ masks.

As for the 4PP test structure fabrication (see Fig. 3(a)), a blanket layer of the material under test was deposited on the textured substrate. In this work we are interested in measuring R$_{sh}$ for front layers that are affected by surface texturing i.e. ITO and Ti/Pd/Ag, taking the solar cell structure we mentioned previously as a reference. Each layer was deposited on top of a nontextured surface, and RPT, dry-NIL and wet-NIL textured surfaces, for comparison. ITO was deposited by sputtering (75 nm) using an in-line sputter system A600 from Leybold Optics, whereas Ti/Pd/Ag were deposited by Pfeiffer PLS 580 electron beam evaporation system (70 nm/50 nm/3 um). 4PP measurements were performed using Automatic Four Point Probe Meter Model 280 from Four Dimensions. As for the TLM test structure depicted in Fig. 3(b), we deposited an ITO layer and then (un)equally spaced contacts of Ti/Pd/Ag with dimensions depicted in Fig. 1. These measurements were conducted using KB-100 system from KB-ESI.

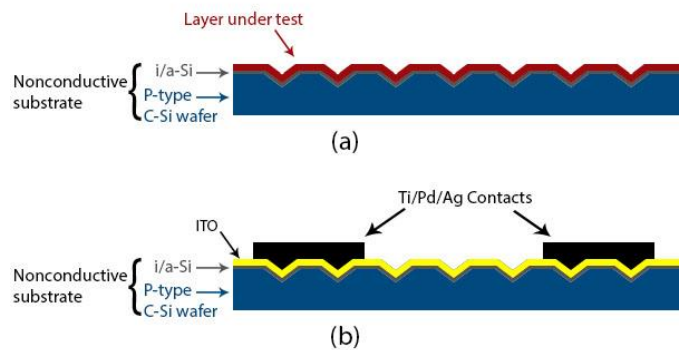

Fig. 3. (a) 4PP test structure with the layer-under-test deposited on a nonconductive textured substrate, and (b) TLM test structure showing Ti/Pd/Ag contacts (only two contacts are shown) deposited on top of an ITO layer. Both test structures are displayed with the wet-NIL texture.

## III. EXPERIMENTAL RESULTS AND DISCUSSION

### A. Topography

Fig. 4(a) to (f) shows cross-sectional SEM images for ITO and Ti/Pd/Ag deposited on top of dry-NIL, wet-NIL and RPT-textured silicon. The RPT are typically 5-10 micron high, while the dry-NIL and wet-NIL are about 500 nm deep. The ITO layer, as one can see in Fig. 4(a), suffers from non-conformal deposition on top of dry-NIL: its thickness is greatly reduced from 75nm at the top flat part to ~10nm at the side walls and the bottom of the dry-NIL. This is because of the presence of a strong slope that prevents the sputtered material to deposit on the side walls. In contrast, ITO has a nearly constant thickness on top of both wet-NIL and RPT (Fig. 4(b) and (c)) thanks to the constant positive slope of pyramids.

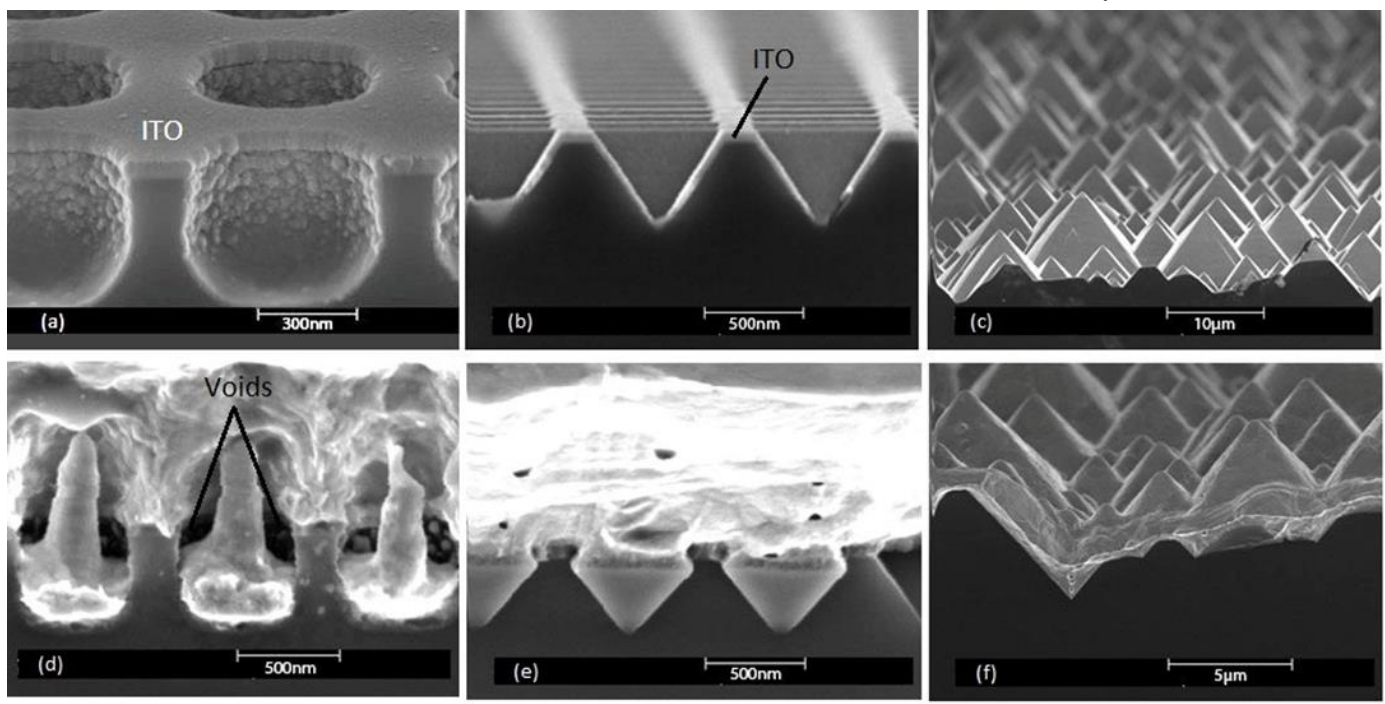

Fig. 4. SEM images for ITO on (a) dry-NIL, (b) wet-NIL and (c) RPT; and Ti/Pd/Ag on (d) dry-NIL, (e) wet-NIL and (f) RPT.

As for Ti/Pd/Ag (Fig. 4(d) to (f)) dry etching prevents the continuous deposition of the layer by introducing voids [Fig. 4(d)] that could be attributed to the combination of two effects: firstly, the high slope of dry-NIL, and secondly, the fact that electron-beam-evaporated Ag can be expected to have a grain size of ~100 nm [22] which is comparable to the dry-NIL feature size. Given its shape, one can suppose this led to shadowing as the layer grows. These voids, however, are absent in the case of wet-NIL where Ti/Pd/Ag is conformally deposited following the shape of the inverted pyramids. As in the case of ITO, RPT also exhibits conformal deposition of the layer deposited on top of it.

### B. TLM

#### 1) Choice of TLM structure

In order to ensure the accuracy of our measurements, we first compared the two TLM structures we introduced in section II.A in terms of their transfer lengths. Both structures are composed of Ti/Pd/Ag contacts deposited on top of an ITO layer on a nontextured silicon wafer, as shown on Fig. 3(b). As shown in Fig. 5 and Table I, we notice that the resulting ITO sheet resistance from the equally spaced structure is lower than that from the unequally spaced one. This is due to the fact that $L_T$ (8.1 µm), calculated using (1), is significantly lower than the length of the metal contacts (L = 1mm). This leads to current flowing into intermediate contacts when the measurement is performed between nonadjacent contacts of the equally-spaced structure [19]. This problem is eliminated in the unequally spaced structure, which is confirmed by the fact that the resulting ITO sheet resistance, using this structure, is consistent with the one measured using 4PP [Fig. 7(a) discussed in section III.C]. We therefore use the unequally-spaced TLM structure in our measurements as it guarantees their accuracy.

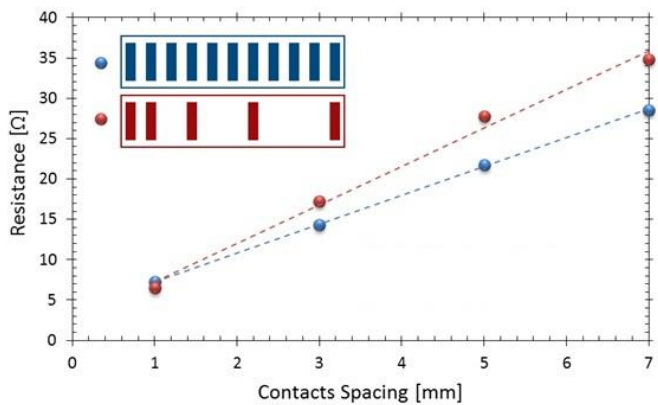

Fig. 5. TLM results comparing the equally and unequally spaced TLM structures. The equally-spaced structure exhibits lower sheet resistance, represented by the line slope, compared to the unequalled-spaced one.

TABLE I
TLM RESULTS COMPARISON BETWEEN THE EQUALLY AND UNEQUALLY-SPACED TLM STRUCTURES. PARAMETERS ARE DERIVED FROM TLM RESULTS DEPICTED IN FIG. 5.

| TLM Structure | $Rsh$ [Ω/□] | $Rc$ [Ω] | $L_T$ [µm] |
|---|---|---|---|
| **Equally-spaced** | 35.6 | 1.87 | 16.6 |
| **Unequally-spaced** | 47.7 | 1.23 | 8.1 |

#### 2) TLM Results

Fig. 6 and Table II show the TLM results of Ti/Pd/Ag contacts deposited on an ITO layer with the substrate textured by RPT, dry-NIL (etching time 45 s) and wet-NIL. One can see that dry-NIL resulted in the highest $R_{sh}$ and $R_c$. RPT resulted in a higher $R_{sh}$ but similar $R_c$ compared to the non-textured substrate. We can suppose that the increased surface area (by ~1.732 times [23][24]) led to thinner layer deposition, both for ITO and metals, increasing their individual sheet resistances, whereas the contact between them has not been significantly affected as in the case of dry-NIL.

On the other hand, wet-NIL resulted in characteristics close to the nontextured case with almost two coinciding TLM lines [see Fig. 6]. $R_{sh}$ has a magnitude close to the nontextured case [see Table II], thanks to conformal deposition on top of the inverted pyramids, as shown by SEM images. However, $R_c$ is significantly higher, although smaller than for dry-NIL.

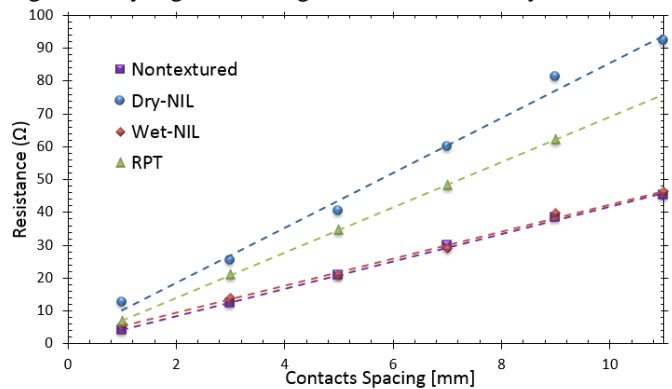

Fig. 6. TLM results of Ti/Pd/Ag contacts on ITO layer deposited on a non-textured and dry-NIL, wet-NIL, and RPT-textured substrates.

TABLE II
TLM RESULTS OF TI/PD/AG CONTACTS ON ITO LAYER DEPOSITED ON A NONTEXTURED AND DRY-NIL, WET-NIL, AND RPT-TEXTURED SUBSTRATES.

| Texture | $Rsh$ [Ω/□] | $Rc$ [Ω] |
|---|---|---|
| **Nontextured** | 41.7 | 0.02 |
| **Dry-NIL** | 83.8 | 0.885 |
| **Wet-NIL** | 41.2 | 0.65 |
| **RPT** | 69 | 0.13 |

### C. Sheet Resistance

In order to further investigate the effect of surface texturing on the front layers, we performed 4PP measurements. As dry-NIL gave the highest sheet resistance, we performed ITO and Ti/Pd/Ag depositions on dry-NIL with varying pattern dimensions. Dry-NIL size was varied by the RIE time, from 0 to 45 seconds, and then $R_{sh}$ was measured for each case. Fig. 7(a) and (b) show $R_{sh}$ of ITO and Ti/Pd/Ag as a function of RIE time of dry-NIL comparing it to RPT and wet-NIL. We notice from Fig. 7(a) that ITO $R_{sh}$ tends to increase as etching time increases, with ~20 Ω/□ increase after 45s etch (deep dry-NIL) compared to the non-textured case (zero RIE time). This is supported by the cross-sectional SEM images in Fig. 4(a) to (c) where ITO is conformally deposited on the inverted pyramids, whereas ITO thickness is greatly reduced at the side walls and the bottom of dry-NIL.

As for the Ti/Pd/Ag, the results are depicted in Fig. 7(b). $R_{sh}$ of Ti/Pd/Ag increases as the etching time increases, with RPT having the highest $R_{sh}$, possibly for the same reason we mentioned in the case of ITO. Wet-NIL has the lowest sheet resistance similar to the nontextured case. These results are supported by the cross-sectional SEM images showed in Fig. 4(d) to (f) depicting the formation of voids inside the Ag layer in the case of dry-NIL only. However, the increase in $R_{sh}$ is in the range of mΩ/□ which is not as significant as in the case of ITO with Ω/□. We can suppose this is owing to the low bulk resistivity of Ag, 1.5 μΩ.cm, compared to 375 μΩ.cm for ITO, calculated from the measured $R_{sh}$ and the deposited thicknesses on nontextured substrates. These measured resistivity values are consistent with values in literature [22][25].

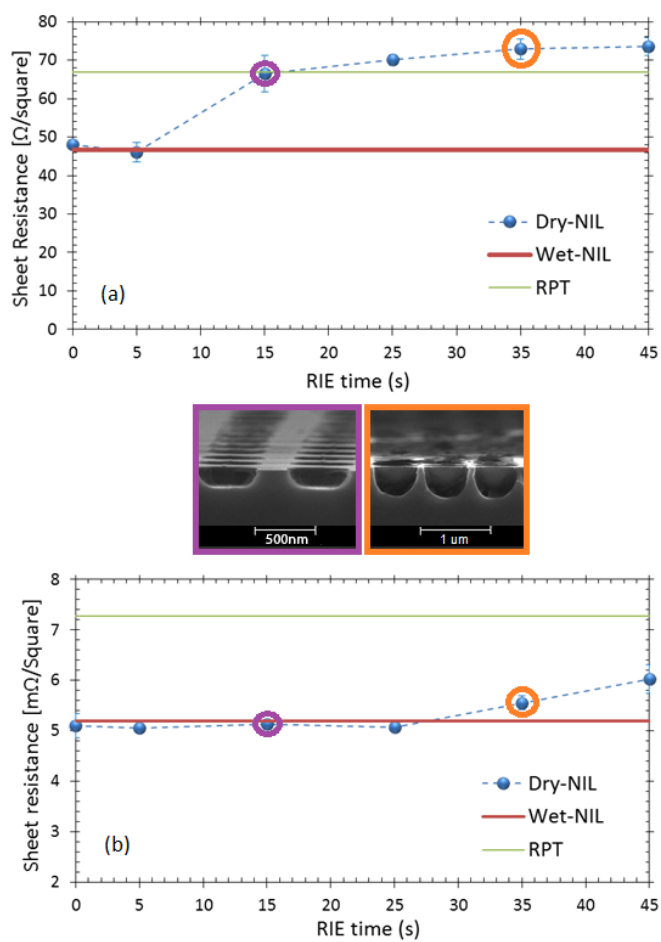

Fig. 7. The effect of dry-NIL size, represented by the RIE time, on the measured sheet resistance of (a) ITO and (b) Ti/Pd/Ag..The results are compared to wet-NIL and RPT with the line width representing the standard deviation of the sheet resistance. The inset shows cross-sectional SEM images that show the dry-NIL pattern shape evolution on a c-Si substrate at 15 s and 35 s RIE time.

## IV. CONCLUSION

Resistance and topographical characterizations have been conducted for solar cells' front layers deposited on top of three different surface topographies, dry-NIL, wet-NIL and RPT. Sheet resistance measurements were realized using the 4PP method for ITO and Ti/Pd/Ag. Contact resistance was measured using the TLM method for Ti/Pd/Ag contacts on top of an ITO layer.

Dry-NIL leads to patterns that increase the sheet resistance of sputtered ITO, which we could attribute to nonconformal deposition on the bottom and sidewalls of the pattern, as revealed by SEM. Moreover, for Ti/Pd/Ag, dry-NIL resulted in voids inside the metal, slightly increasing its sheet resistance. For the same combined reasons, contact resistance increased as well. On the other hand, wet-NIL resulted in a sheet resistance with a magnitude close to the nontextured case for all layers. This was attributed to its smooth slopes along the crystalline axes, characteristic of the alkaline wet etching of a (100) c-Si surface.

Such measurements give a good insight of how surface texturing could affect the FF of c-Si HJ solar cells. TCO has been found to be a sensitive material to surface texturing due to its high resistivity compared to metals. Because of this sensitivity, the shape of the nanopatterns should be carefully tuned to keep a good conformality of the TCO.